\documentclass[12pt]{article}
\usepackage[letterpaper]{geometry}
\usepackage{amssymb,amsmath,amsthm}
\usepackage{graphicx}
\usepackage{natbib} 
\usepackage{tikz}
\usetikzlibrary{shapes,shadows,arrows}
\usepackage{caption}
\usepackage{bm}
           
\newcommand{\med}{\mathop{\mbox{med}}}
\newcommand{\AO}{\text{AO}}

\begin{document}

\title{Finding Outliers in Surface Data and Video}

\author{Mia Hubert, Jakob Raymaekers, Peter J. Rousseeuw,\\
        Pieter Segaert\footnote{
        Department of Mathematics, KU Leuven,
		    Belgium}~\footnote{
        email: Mia.Hubert@wis.kuleuven.be}}
					
\date{January 29, 2016}

\maketitle

\begin{abstract}
Surface, image and video data can be considered as functional data with a bivariate domain.
To detect outlying surfaces or images, a new method is proposed based on the mean and the variability of the degree of outlyingness at each grid point. A rule is constructed to flag the outliers in the resulting functional outlier map. Heatmaps of their outlyingness indicate the regions which are most deviating from the regular surfaces. The method is applied to fluorescence excitation-emission spectra after fitting a PARAFAC model, to MRI image data which are augmented with their gradients, and to video surveillance data. 
\end{abstract}


\section{Introduction}
\label{sec:introduction}
The most common type of functional data are curves, typically measured over time. Also spectral data, measured over a range of wavelengths, can be considered as functional data and can be analyzed as such~\citep{Saeys:Functional}. When several measurements are taken at each time point, we obtain multivariate functional data. Bivariate examples are height and weight curves of children, temperature and dewpoint temperature at several weather stations during several days, or measurements of the
human heart activity at two different places on the body \citep{Claeskens:MFHD}. Whereas the domain of these data is always univariate, the response can thus be multivariate.

Surface data are an increasingly common data type in several fields of research. In chemistry for example, fluorescence spectroscopy is a well known technique that yields surface data with an excitation and an emission dimension. In geography, surface data is obtained by measuring characteristics such as daily precipitation on a certain area of land. 
Also digital image and video data are represented by a grid of pixels and typically contain grayscale values or three-dimensional RGB values that define the color intensities. 
All these advanced data types can thus be seen as multivariate functional data with a bivariate domain. 

To detect outlying surfaces or to flag outlying parts of a surface, it is well known that classical statistical techniques are not trustworthy as the analysis itself may be distorted by the outliers.
Outlier detection for multivariate functional data with a univariate domain has been studied in depth in \cite{Hubert:MFOD}. 
In that paper a taxonomy of outlying curves was proposed, and distribution-free statistical methods were introduced to measure 
their degree of outlyingness.  
\cite{Hubert:MFOD_Rejoinder} illustrated this on a raw 
excitation-emission fluorescence dataset, and proposed the 
functional outlier map (FOM) as a graphical display to visualize 
the functions according to their degree and type of outlyingness.  

In this paper, we build on these methods with a threefold objective.
First, we propose a cutoff to distinguish between the outlying and 
the regular surfaces, which makes the FOM more informative. 
Secondly, we improve the outlier detection ability for surface data by first modeling the data according to a multiway model and applying the FOM to the residuals instead of the raw data.
Finally, we illustrate the extension of these techniques 
to multivariate image and video data. 

The next section introduces the methodology behind the FOM as well as our new outlier detection rule. 
In Section \ref{sec:surface} we apply the new rule to excitation-emission matrices before and after fitting a 
PARAFAC model to them.
Section~\ref{sec:image} demonstrates the performance of our method on MRI image data and introduces the idea of including
gradients in the analysis.
In Section~\ref{sec:video} we apply the methodology to a 
video consisting of 633 color images. 

\section{Methodology}
\label{sec:Method}

\subsection{Functional Adjusted Outlyingness}

The main concept for measuring the degree of outlyingness 
is the \textit{adjusted outlyingness} (AO) 
of~\cite{Brys:RobICA}. 
Its construction is recalled in the Appendix.
For univariate data, the AO is a robust version of the 
absolute $z$-score as it measures an observation's 
deviation from the median. 
It allows for skewness in the data by estimating scale
separately on either side of the median. 
For multivariate data, the AO of a point is defined
as its maximal univariate AO when projecting the data
on many directions.  

To extend the AO to multivariate functional data, 
\cite{Hubert:MFOD} defined the \textit{functional adjusted
outlyingness} (fAO) of a $p$-dimensional curve $X$ as 
the (weighted) average of its AO values over all points
of its domain.
More precisely, let\linebreak 
$\bm{Y}=\{Y_1,Y_2,\ldots,Y_n\}$ be a 
sample of $p$-dimensional curves, recorded at time 
points $\{t_1,\ldots,t_T\}$. 
The fAO of $X$ with respect to $\bm{Y}$ is then defined as
\begin{equation} \label{eq:fao}
  \text{fAO}(X; \bm{Y}) = \sum_{j=1}^{T}{\AO(X(t_j);
	\bm{Y}(t_j)) \; W(t_j) }
\end{equation}
where $\bm{Y}(t_j)$ is a sample in $p$ dimensions and
$W(.)$ is a weight function for which 
$\sum_{j=1}^{T}{W(t_j)}=1$. 
This weight function allows to assign a different importance 
to the outlyingness of a curve at different time points. 
One could for example downweight time points near the 
boundaries if measurements are recorded less precisely at 
the beginning and the end of the process.   

Definition~\eqref{eq:fao} for functional data with a univariate domain can easily be extended to functions with a bivariate domain such as surfaces and images. We then assume that the outcome of the experiment is recorded at a grid of discrete points, e.g.\ discrete excitation and emission wavelengths. 
Therefore, it is convenient to use two indices $j=1,\ldots,J$ and $k=1,\ldots,K$, one for each dimension of the grid, to characterize these points. A surface or image can then be represented by a $J \times K$ matrix where each cell $(j,k)$ contains the height of the surface or the color intensity at that grid point.
We then define the functional adjusted outlyingness of
a multivariate function $X$ with bivariate domain relative 
to a dataset $\bm{Y}$ of functions recorded at grid points 
$\{(j,k) ; j=1,\ldots, J \mbox { and } k=1,\ldots, K\}$ as
\begin{equation}\label{eq:fao_surface}
\text{fAO}(X; \bm{Y}) = \sum_{j=1}^{J}{\sum_{k=1}^{K}{\AO(X(j,k);\bm{Y}(j,k)) \; W_{jk}}}
\end{equation}
with $\sum_{j=1}^J \sum_{k=1}^K W_{jk} = 1$.
For illustrative purposes we will use a uniform weight
function throughout this paper, i.e. $W_{jk}=1/(JK)$ for 
all $j=1,\ldots,J$ and $k=1,\ldots,K$. 
Definition~\eqref{eq:fao_surface} can trivially be extended
to functions with a multivariate domain of dimension $d>2$,
such as three-dimensional images consisting of voxels.

The fAO can be applied to raw data, but in other applications
it might be interesting to first fit a parametric model to the data, after which the fAO can be computed on the residuals. 
For multiway data, PARAFAC is a commonly used method to fit a trilinear model. The concept of applying fAO to the residuals of a PARAFAC model is illustrated on a real data example in Section~\ref{sec:surface}.

When analyzing curves, it is often informative to consider 
the derivative of the curves as well. 
In particular, this is helpful when the goal is to detect
curves with a deviating shape, see \cite{Hubert:MFOD} for 
some examples. 
For surfaces or images, we can augment the available raw 
data by including gradients instead.
This idea will be illustrated in Section~\ref{sec:image}.

\subsection{Functional outlier map}
The functional outlier map (FOM) was introduced 
in~\citep{Hubert:MFOD_Rejoinder} as a graphical tool to
visualize the degree of outlyingness of curves. A similar definition can be applied to surfaces or images. 
The FOM is a scatter plot which displays each function's 
fAO on the horizontal axis and vAO, a measure of the 
variability of its AO over all points of the domain, on 
the vertical axis. 
More precisely, for each $Y_i$ the vAO is defined as
\begin{equation}\label{eq:vAO}
\text{vAO}(Y_i;\bm{Y})=\text{stdev}_{j,k}(\text{AO}(Y_i(j,k);\bm{Y}(j,k)))/(1+\text{fAO}(Y_i;\bm{Y}))
\end{equation}
so the FOM is a scatter plot of the points
\begin{equation}\label{eq:fom}
\left(\;\text{fAO}(Y_i;\bm{Y}),\text{vAO}(Y_i;\bm{Y})\;\right) 
\end{equation}
for $i=1,\ldots,n$.
We divide by fAO in Equation~\eqref{eq:vAO} in order to 
measure relative instead of absolute variability. 
This can be understood as follows. 
Suppose that the functions are centered around zero and 
that $Y_l(j,k)=2 \; Y_i(j,k)$ for all $j$ and $k$. 
Then $\text{stdev}_{j,k}(\text{AO}(Y_l(j,k);\bm{Y}(j,k)))=2 \; \text{stdev}_{j,k}(\text{AO}(Y_i(j,k);\bm{Y}(j,k)))$ but their relative variability is the same. 
Because $\text{fAO}(Y_l;\bm{Y})=2\;\text{fAO}(Y_i;\bm{Y})$, dividing by fAO solves this problem.

The objective of the FOM is to reveal outliers in the data, 
and its interpretation is fairly straightforward. 
We use the taxonomy of functional outliers presented 
by~\cite{Hubert:MFOD} to distinguish between different types of 
outliers. 
Points in the lower left part of the FOM represent regular 
surfaces which hold a central position in the dataset. 
Points in the lower right part are surfaces with a high fAO 
but a low variability of their AO values. 
This happens for shift outliers, i.e.\ surfaces which have the 
same shape as the majority but are shifted on the whole domain. 
Points in the upper left part have a low fAO but a high vAO. 
Typical examples are isolated outliers, i.e.\ surfaces which 
only display outlyingness over a small part of their domain.
The points in the upper right part of the FOM have both a high 
fAO and a high vAO. These correspond to surfaces which are 
strongly outlying on a substantial part of their domain.

In this paper we add an additional feature to the FOM, namely 
a rule to flag the outliers. 
For this purpose we define a summary measure, the 
\textit{combined functional outlyingness} (CFO) of a 
surface $Y_i$ as
\begin{equation} \label{eq:cfo}
  \text{CFO}_i=\text{CFO}(Y_i; \bm{Y}) = \sqrt{(\text{fAO}_i
	/\med(\text{fAO}))^2+(\text{vAO}_i/\med(\text{vAO}))^2}
\end{equation}
where $\text{fAO}_i = \text{fAO}(Y_i;\bm{Y})$,  
$\med(\text{fAO}) = \text{median}(\text{fAO}_1,\ldots,\text{fAO}_n)$ 
and similarly for\linebreak vAO. 
Note that the CFO characterizes the points on the FOM through 
their Euclidean distance to the origin, after scaling. 
We expect outliers to have a large CFO. 
In general, the distribution of the CFO is unknown but 
skewed to the right. 
To define a cutoff, we first symmetrize the CFO by the logarithmic 
transformation. 
Then we center and scale the resulting values in a robust way and 
compare them with a high quantile of the gaussian distribution. 
More precisely,  let $\text{LCFO}_i = \log (0.1 + \text{CFO}_i)$ 
for all $i=1,\ldots,n$. Then we flag function $Y_i$ as an outlier if
\begin{equation} \label{eq:cutoff}
  \frac{\text{LCFO}_i-\med(\text{LCFO})}
	{\text{MAD}(\text{LCFO})} > \Phi^{-1}(0.995) \, .
\end{equation}
In the denominator, MAD denotes the median absolute 
deviation defined as\linebreak
$\text{MAD}(\text{LCFO})=1.4826 \; \med(| \; \text{LCFO}-\med(\text{LCFO})|)$.

In addition to the FOM, it is often instructive to plot 
the AO values themselves over their domain. 
For a function with a bivariate domain, this yields a
two-dimensional heatmap. 
These plots can reveal particularly outlying regions and provide 
more insight into why certain surfaces are flagged as outliers 
by the FOM. 
This will be illustrated in the following sections.

\section{Surface data}\label{sec:surface}

\begin{figure}[!htb]
\centering
\includegraphics[width=1.0\textwidth]{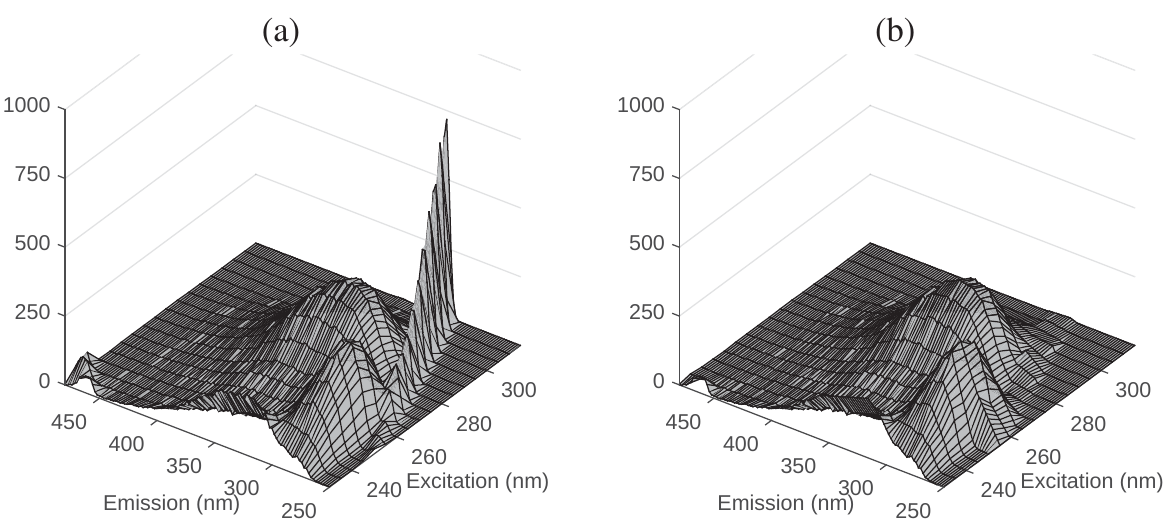}
\caption{(a) Original landscape 8 with scatter and (b) with scatter removed.}
		\label{fig:L8}
\end{figure}

The Dorrit dataset contains excitation-emission landscapes of 
27 mixtures of 4 fluorophores in an aqueous solution and has been studied extensively by \cite{Engelen:Scatter}, \cite{Engelen:robParafac}  and \cite{Hubert:RParafac-SI}. Such landscapes are typically stored in excitation-emission matrices (EEM). These matrices contain measurements of fluorescence intensity for 116 emission spectra ranging from 250 to 482 nm at 18 excitation wavelengths ranging from 230 to 315 nm. 
This yields 27 samples $Y_i$, each containing measurements 
$Y_i(j,k)$ for $j=1,\ldots,J=18$ and $k=1,\ldots,K=116$.
Raman and Rayleigh scattering, showing up in 
Figure~\ref{fig:L8}(a) as two diagonal ridges in 
landscape 8, are a common problem in this type of experiment
and can severely distort the analysis.
To resolve this we used the method of \cite{Engelen:Scatter} 
which automatically identifies the scattering. 
After setting the scattering to missing values, we imputed
them by interpolating each excitation profile.
The result of this process is illustrated in 
Figure~\ref{fig:L8}(b).

\begin{figure}[!htb]
\centering
 \centering
    \includegraphics[width=0.6\textwidth]{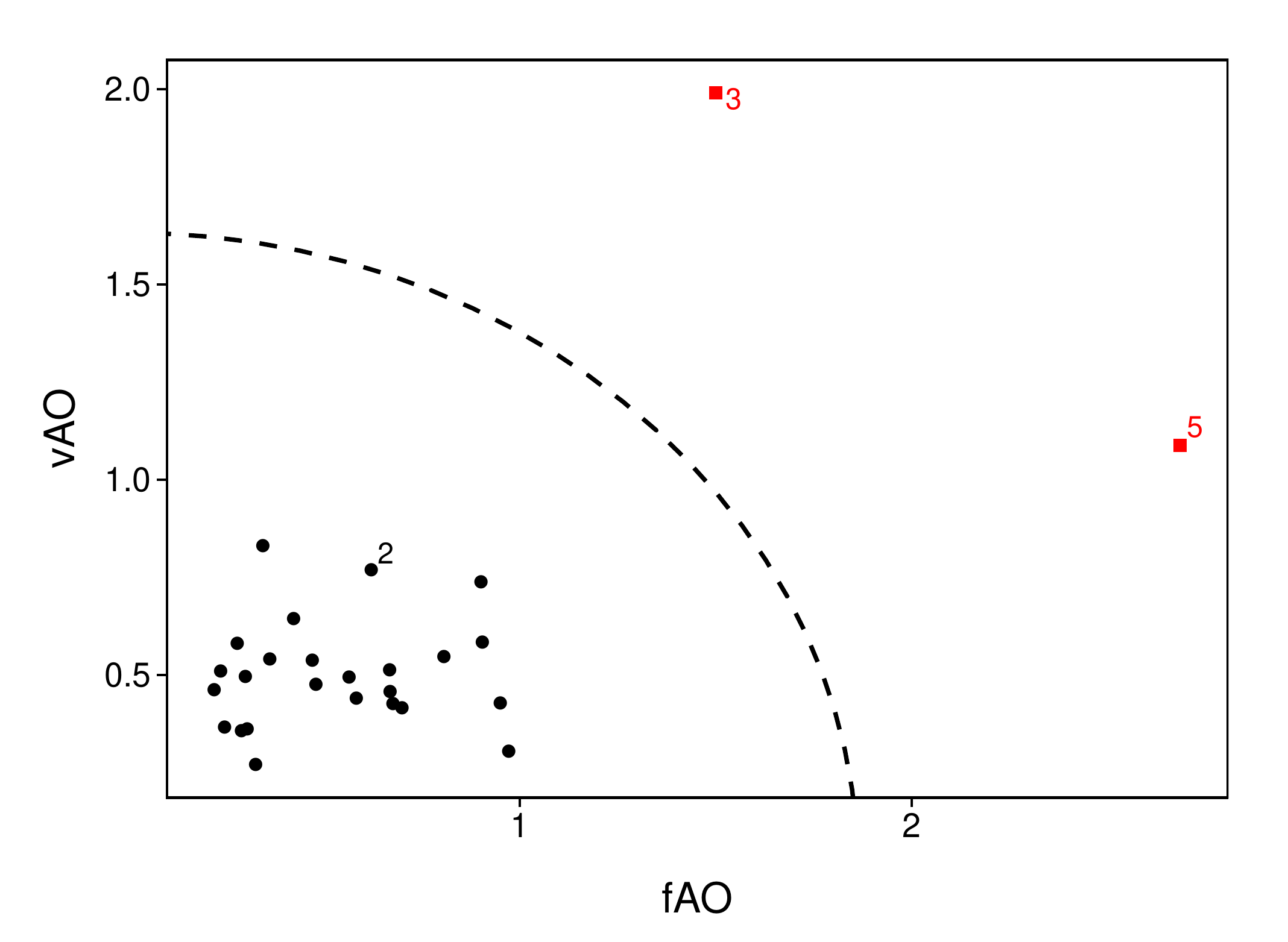} 
    \caption{FOM of the raw Dorrit data.}
    \label{fig:dorritfom}
\end{figure}

Using Definition~\eqref{eq:fao_surface} with a uniform 
weight function, we obtain the functional outlier map 
depicted in Figure~\ref{fig:dorritfom}. 
The dashed black line is determined by the cutoff 
value~\eqref{eq:cutoff} and shows the boundary between 
the regular data and the outliers.
The outliers, plotted as red squares, are landscapes
3 and 5.

We can improve our analysis by incorporating more 
information, for example by applying parallel factor 
analysis (PARAFAC) to the data. PARAFAC is known to be a powerful tool for analyzing multiway data and can in particular be used to decompose three-way EEM data into a trilinear model. This trilinear model states that the value of a landscape $Y_i$ at EE-point $(j,k)$ can be written as the product of three corresponding values of the score matrix $A$ and the loading matrices $B$ and $C$, plus an error term:
\begin{equation}\label{eq:parafac}
Y_{i}(j,k)=\sum_{f=1}^{F}{a_{if}b_{jf}c_{kf}}+e_{ijk} \;.
\end{equation}
Here $F$ denotes the number of components which we take as $F=4$, corresponding to the 4 known fluorophores present in the mixtures. 
Applying PARAFAC to the data results in 27 residual 
landscapes $\widetilde{Y}_i(j,k)=e_{ijk}$. 
With the objective of outlier detection in mind, it is important to use a robust PARAFAC algorithm as it is well known that the presence of outliers can severely impact the PARAFAC decomposition. We have used the robust PARAFAC algorithm proposed by \cite{Engelen:robParafac} in which we set the robustness parameter $h=0.75$. This  parameter indicates that the PARAFAC model is based on the $75\%$ of surfaces which yield the smallest residuals. 

\begin{figure}[!htb]
\centering
\includegraphics[width=0.6\textwidth]{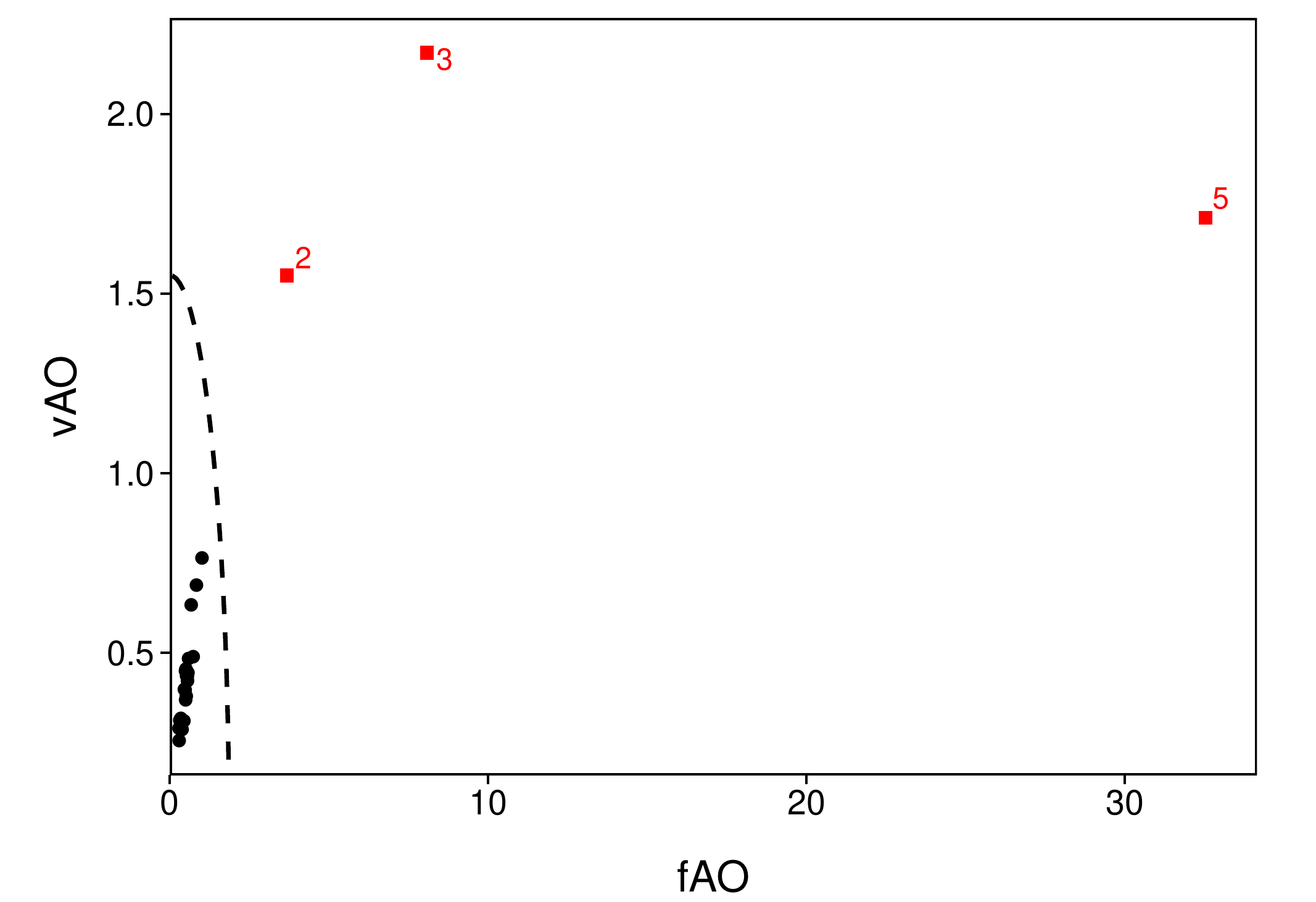} 
\caption{FOM of the residuals of the Dorrit data.}
\label{fig:dorritfom_parafac}
\end{figure}

By computing the fAO on the residuals instead of the original data, the trilinear structure of the dataset is taken into account. 
The resulting functional outlier map is shown in 
Figure~\ref{fig:dorritfom_parafac} and flags 
landscapes 2, 3 and 5.
These outliers were also detected 
in~\cite{Engelen:robParafac}.
Note that points 3 and 5 lie much further from the boundary 
(the dashed black curve) than in the FOM of 
Figure~\ref{fig:dorritfom}. 
This indicates that surfaces 3 and 5 have outlying values but mostly expose an outlying structure. 
Landscape 2 barely contains abnormally low or high values but 
has large residuals with respect to the PARAFAC model fitted
to the majority of EEM landscapes. 

\begin{figure}[!htb]
\centering
\includegraphics[width=1.0\textwidth]{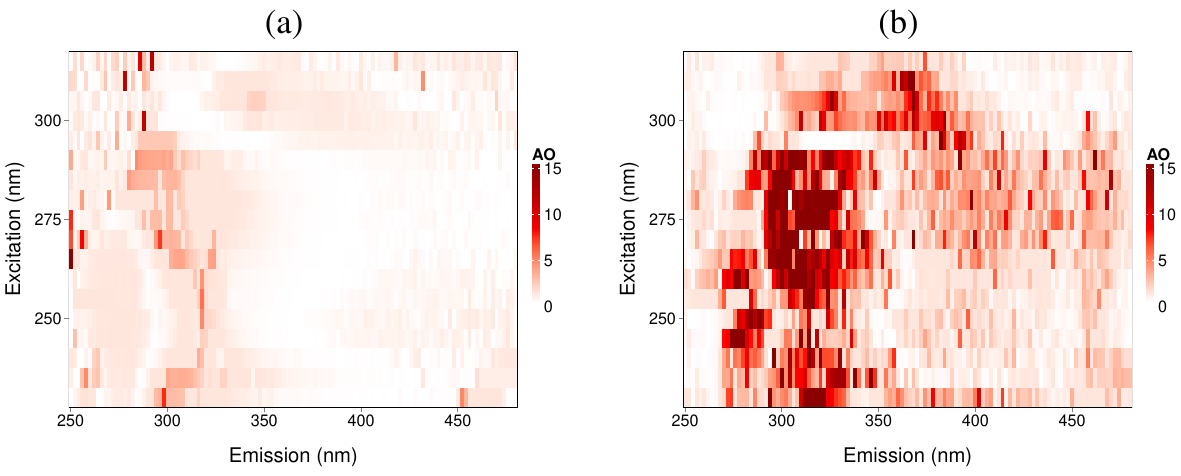}
\caption{Heatmap of the AO values of landscape 2 for 
         (a) the raw values, 
         (b) the residuals after a robust PARAFAC fit.}
		\label{fig:AOmap_dorrit}
\end{figure}

A heatmap of the AO values over the domain allows us to 
inspect surface 2 in more detail. 
Figure \ref{fig:AOmap_dorrit}(a) plots the AOs of the 
original landscape 2, whereas 
Figure \ref{fig:AOmap_dorrit}(b) shows the AOs of its 
PARAFAC residuals. 
To improve visibility, all AO values of 15 or higher 
received the darkest color.
The left plot indicates that landscape 2 indeed does not 
have large outlying regions, as the vast majority of its
cells are light-colored. 
However, the right plot shows that its residuals are significantly outlying in the emission spectra between 
275 and 350 nm. 
This indicates that the shape of landscape 2 differs 
strongly from the shape predicted by the PARAFAC model. Therefore, this landscape can be classified as a shape 
outlier.

\section{Image data}\label{sec:image}

The MRI dataset contains brain imaging data of 416 persons, aged between 
18 and 96 years~\citep{Marcus:OASIS} and can be freely accessed 
at {\it www.oasis-brains.org}.
For each person, multiple MRI scans are available including 3 to 4 raw MRI 
images and several processed scans such as an averaged image of the raw 
scans, the atlas-registered gain field-corrected image and a masked version
 of the latter scan resampled to 1mm isotropic pixels. 
Masking sets all non-brain pixels to an intensity value of zero. 
Moreover these images are normalized, meaning that the size of the head is 
exactly the same in each image.  
The masked images have 176 by 208 pixels with grayscale values between 0 
and 255. 
All together we thus have 416 observed images $Y_i$ containing univariate 
intensity values $Y_i(j,k)$, where $j=1,\ldots,J=176$ and $k=1,\ldots,K=208$.

There is more information in the image than just the raw values. 
We can incorporate part of the shape information by computing the gradient 
in every pixel of the image. 
This gradient is composed of the derivatives in the horizontal and the 
vertical direction. 
More precisely, the gradient in pixel $(j,k)$ is defined as the 2-dimensional 
vector $\nabla Y_i(j,k) = \left(\frac{\delta Y_i(j,k)}{\delta j},
\frac{\delta Y_i(j,k)}{\delta k} \right)$. 
In practice we do not have the exact derivatives of the image at the 
recorded pixels, so we have to rely on numerical approximations. 
We use forward and backward finite differences to approximate the 
derivatives in the pixels at the boundary of the brain. 
For the other pixels, we have used central differences.
In the horizontal direction these are given by
\begin{align*}
\frac{\delta Y_i(j,k)}{\delta j} \approx &\frac{-3 \, Y_i(j,k) + 4 \, Y_i(j+1,k) - Y_i(j+2,k)}{2} &&(\text{forward difference})\\
\frac{\delta Y_i(j,k)}{\delta j} \approx &\frac{Y_i(j+1,k)-Y_i(j-1,k)}{2}  &&(\text{central difference})\\
\frac{\delta Y_i(j,k)}{\delta j} \approx &\frac{Y_i(j-2,k)-4 \, Y_i(j-1,k)+3\,  Y_i(j,k)}{2}  &&(\text{backward difference}).
\end{align*}
The derivatives in the vertical direction are computed analogously.

\begin{figure}[!htb]
\centering
\includegraphics[width=1\textwidth]{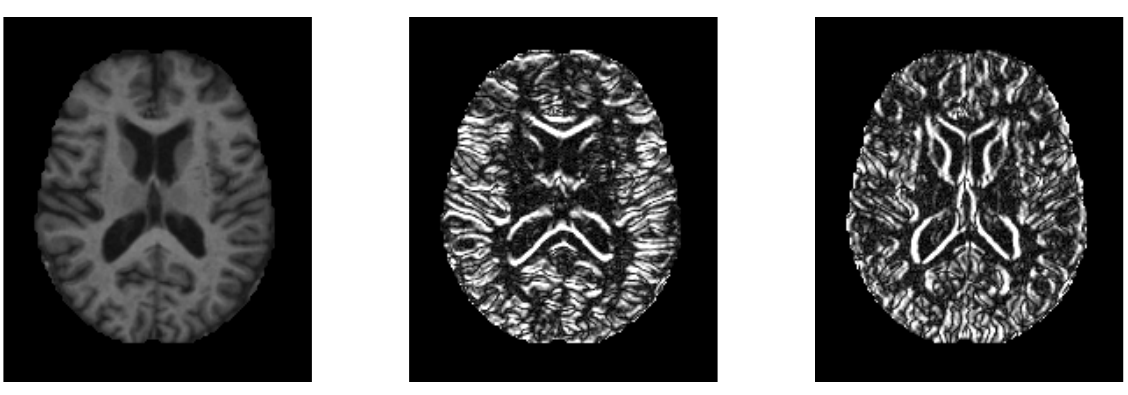}
\caption{Original MRI image of person 144 and its derivatives in the 
         horizontal and vertical direction.}
\label{fig:mridata}
\end{figure}

Adding these derivatives to the data yields a multiway dataset of dimensions 
$416 \times 176 \times 208 \times 3$, so each $Y_i(j,k)$ is trivariate. 
For each person we thus have three data matrices which represent the original 
MRI image and its derivatives in both directions. 
Figure~\ref{fig:mridata} represents these three matrices for person 144.

\begin{figure}[!htb]
\centering
\includegraphics[width=0.6\textwidth]{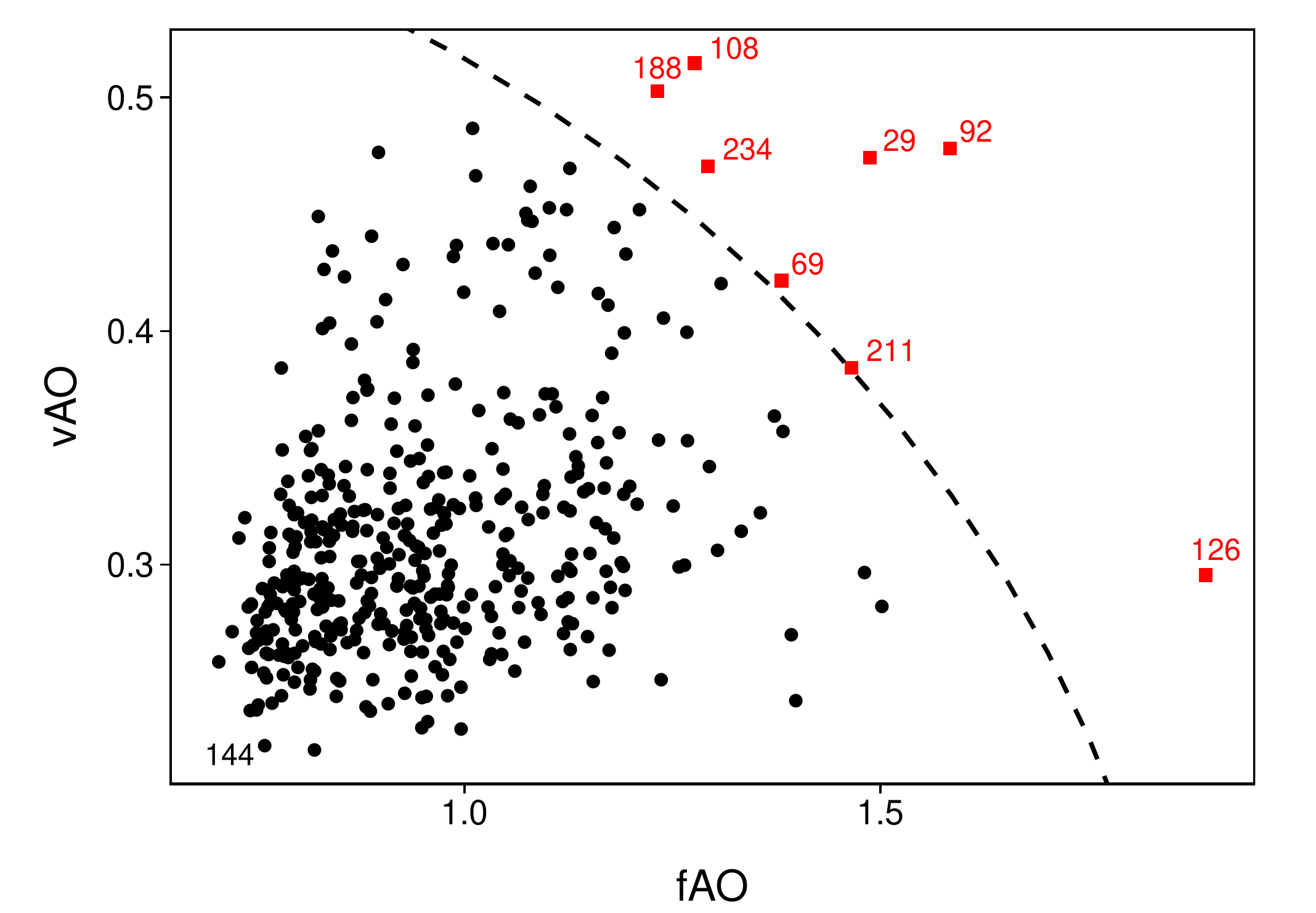} 
\caption{FOM of the MRI images dataset.}
\label{fig:mrifom}
\end{figure}

The functional adjusted outlyingness of an MRI image $Y_i$ relative to the 
sample $\bm{Y}$ is given by~\eqref{eq:fao_surface} :
\begin{equation}
\text{fAO}(Y_i; \bm{Y}) = \frac{1}{176 \times 208} \sum_{j=1}^{176}
                          \sum_{k=1}^{208} \AO(Y_i(j,k); \bm{Y}(j,k)) \; ,
\end{equation}
where $\AO(Y_i(j,k);\bm{Y}(j,k))$ is the adjusted outlyingness of the
trivariate pixel $(j,k)$. 
The resulting FOM is presented in Figure~\ref{fig:mrifom}.

This plot indicates the presence of 8 outliers of several types. 
Image 126 has a remarkably high fAO combined with a relatively low vAO. 
This suggests a shift outlier, i.e.\ a function whose values are all
shifted relative to the majority of the data. 
Images 29 and 92 have a large fAO in combination with a high vAO, 
indicating that they have strongly outlying subdomains. 
Images 108, 188 and 234 have an fAO which is on the high end relative 
to the dataset but which by itself does not make them outlying. 
Only in combination with their large vAO are they flagged as outliers. 
Images 69 and 211 have a CFO value just slightly above the 
cutoff, meaning they are borderline cases.

\begin{figure}[!htb]
\centering
%
%
\includegraphics[width=0.81\textwidth]{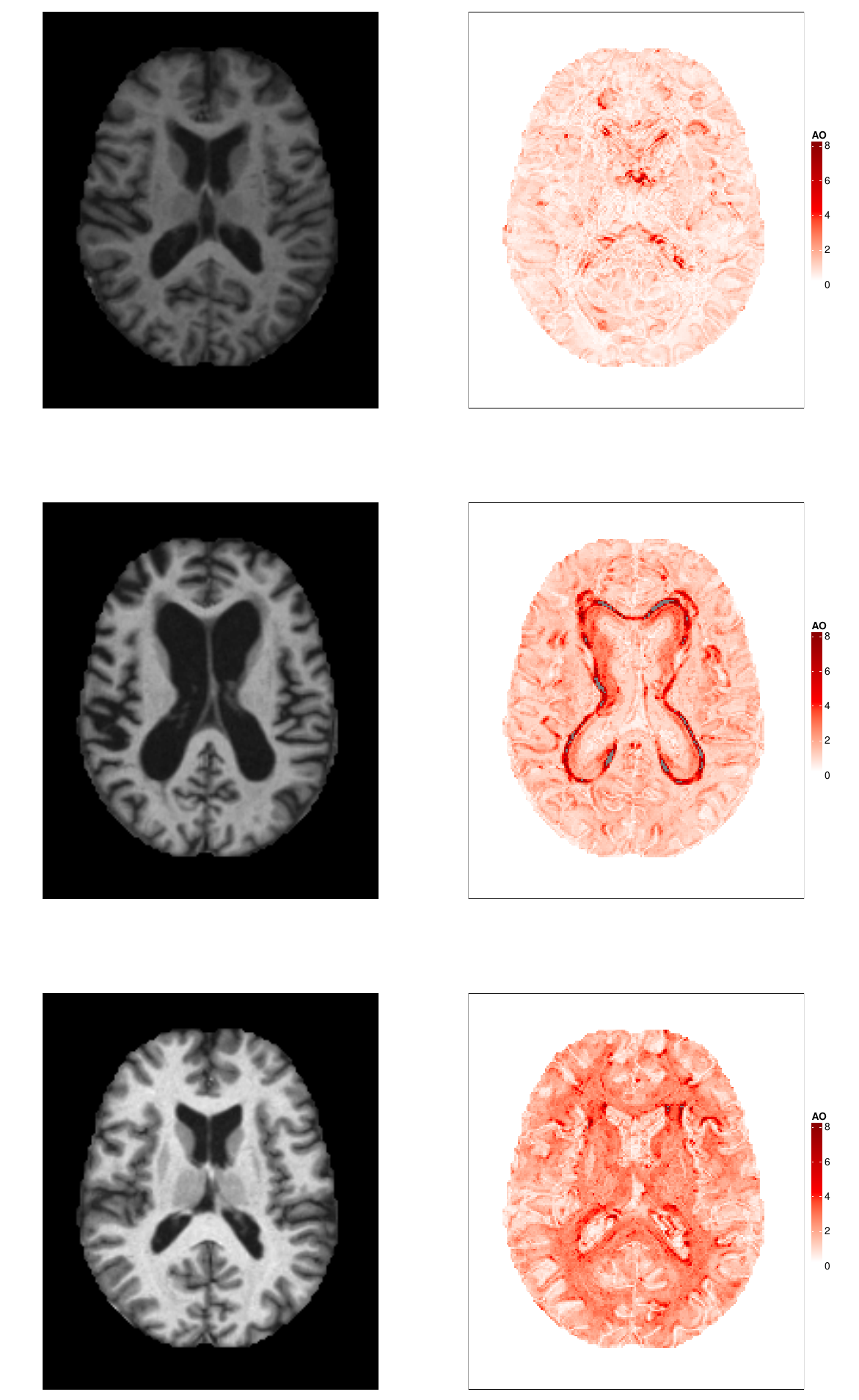}
\captionsetup{format=hang}
\caption{MRI image (left) and AO heatmap (right) of persons 144 (top), 
         92 (middle) and 126 (bottom).}
\label{fig:mrigrid}
\end{figure}

A heatmap of the AO values can help to understand the root cause of 
an image's outlyingness.
In Figure~\ref{fig:mrigrid} we compare the MRI images (left) and the 
AO maps (right) of persons 144, 92, and 126. 
(AO values of 8 or higher received the darkest color.)
Image 144 has the smallest CFO value, and as such can be thought of as
the least outlying image in the dataset. 
As expected, the AO heatmap of image 144 shows very few outlying 
pixels.
For person 92, the AO heatmap nicely marks the region in which the MRI 
image deviates most from the majority of the images. 
Note that the boundaries of this region have the highest outlyingness. 
This is the result of including the derivatives in the analysis, as 
they emphasize the pixels at which the grayscale intensity changes. 
The AO heatmap of person 126 does not show any extremely outlying 
region but a rather high outlyingness over the whole domain, 
which explains its large fAO and regular vAO value. 
The actual MRI image to its left is globally lighter than the 
others, which confirms that it is a shift outlier.

\section{Video data}\label{sec:video}

\begin{figure}[!htb]
\centering
\includegraphics[width=1\textwidth]{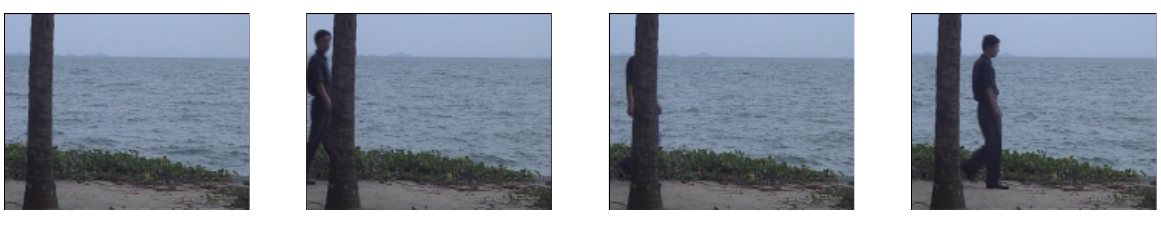}
\caption{Frames number 100, 487, 491 and 500 from the video dataset.}
\label{fig:videodata}
\end{figure}

This dataset consists of 633 images which together form a surveillance 
video of a beach, filmed with a static camera~\citep{Li:ObjectDetection}. 
The data and the original video can be found
at {\it http://perception.i2r.a-star.edu.sg/bk\_model/bk\_index.html}. 
The video first shows a beach with a tree during 8 seconds, shown in the 
leftmost frame of Figure~\ref{fig:videodata}. 
Then a man enters the screen from the left (second frame), disappears 
behind the tree (third frame), and then reappears again on the right 
side of the tree and stays on screen until the end of the video. 
The aim of applying our method to this dataset is to detect the man 
in this video. 
The images have $160 \times 128$ pixels and are stored using the RGB 
(Red, Green and Blue) color model, so each image corresponds to three 
matrices, each of dimensions $160 \times 128$. 
Overall we have 633 images $Y_i$ containing trivariate $Y_i(j,k)$ 
for $j=1,\ldots,J=160$ and $k=1,\ldots,K=128$. 

\begin{figure}[!htb]
\centering
\centering
\includegraphics[width=0.6\textwidth]{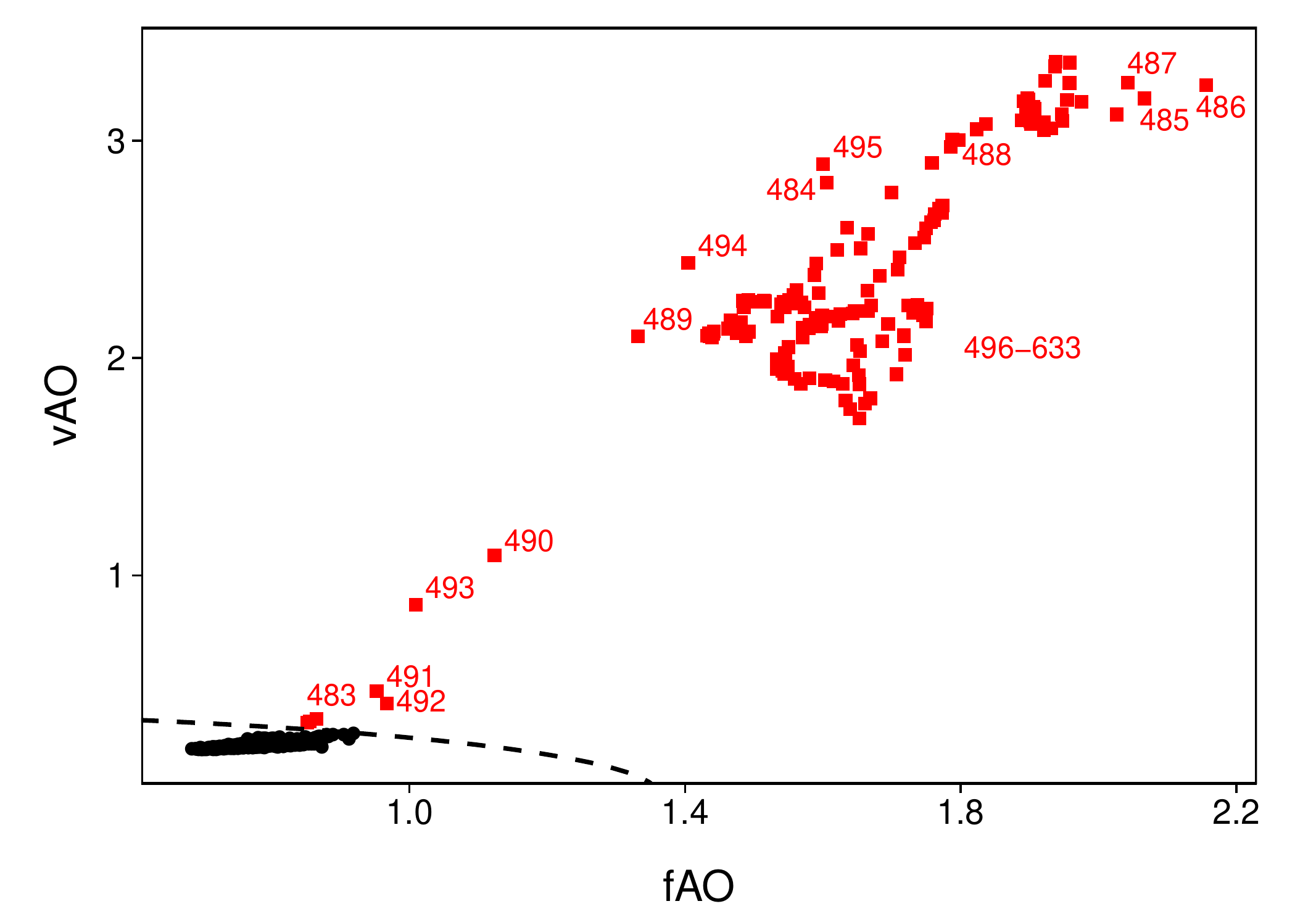} 
\caption{FOM of the video dataset.}
\label{fig:videofom}
\end{figure}

Computing the fAO according to~\eqref{eq:fao_surface} yields the FOM 
in Figure~\ref{fig:videofom}. 
It is very instructive as the path of the man in the video can be traced 
in it.
The first 480 images, which depict the beach and the tree and where only 
the water surface causes slight variations, are found on the bottom left 
side of the FOM inside the dashed curve that separates the regular 
frames from the outliers.
Around frame 483 the man enters and as a result the standard deviation 
of the adjusted outlyingness rises slightly. 
The fAO itself increases more slowly as the fraction of the pixels 
covered by the man is still low. 
These frames can thus be classified as isolated outliers. 
Their CFO barely exceeds the cutoff~\eqref{eq:cutoff} so on the FOM they 
are still close to the dashed line. 
Frames 484--487 have a very high fAO and vAO. 
In these images, the man is clearly visible between the left border of 
the frame and the tree. 
Consequently these images have outlying pixels in a substantial part 
of their domain. 
Frames 488--491 see the man disappear behind the tree. 
In these frames the fAO goes down as the fraction of outlying
pixels decreases.
From frame 492 onward the man reappears on the right side of the tree 
and stays in the screen until the end, as is reflected in 
the FOM since these frames contain many outlying pixels.

\begin{figure}[!htb]
\centering
\includegraphics[width=1.0\textwidth]{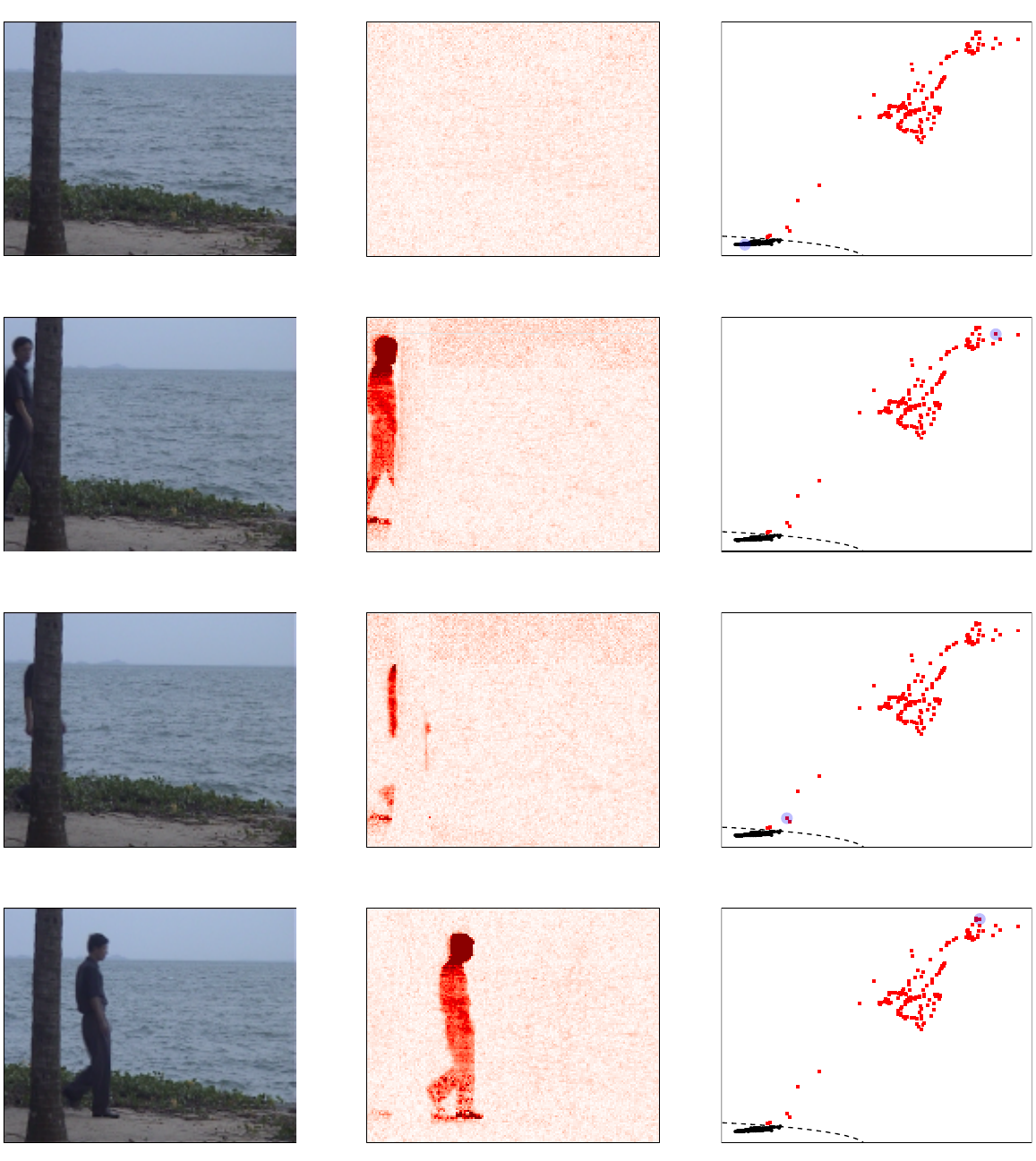}
\captionsetup{format=hang}
\caption{Frames 100, 487, 491 and 500 from the video dataset 
         (left), the corresponding AO heatmaps (middle), and
				 the FOM with marker on the position of the 
				 corresponding frame (right).}
\label{fig:videosum}
\end{figure}

In addition to the FOM we can construct AO heatmaps of individual 
frames.  
For frames 100, 487, 491 and 500, Figure~\ref{fig:videosum} shows the 
raw image on the left, the AO heatmap in the middle and the FOM on 
the right. 
On the FOM we have drawn a blue circle around the relevant frame to 
indicate its location. 
This sequence of plots clearly shows that the proposed method works 
very well for this surveillance video data: not only can the man's 
path be followed on the FOM, it is also clear from the AO heatmaps 
where exactly the man is in those frames.    
We have created a video in which the raw image, the AO heatmap and
the FOM evolve over time alongside each other. 
It can be downloaded from {\it http://wis.kuleuven.be/stat/robust/publ}.

%

\section*{Acknowledgment} 

MH, JR and PR acknowledge the financial support of grants C16/15/068 
and\linebreak STRT/13/006 of the Internal Fund KU Leuven.

\vskip0.2in	

\bibliographystyle{apalike} 
%

\vskip1cm

\Large
\noindent{\bf Appendix}\\ 
\normalsize

\vskip0.3cm

The adjusted outlyingness of multivariate data has been introduced
by~\cite{Brys:RobICA} and was studied in detail 
by~\cite{Hubert:OutlierSkewed}. 
We first recall the adjusted outlyingness of a point $x$ relative to 
a univariate dataset $Y=\{y_1,\ldots,y_n\}$. 
It can be seen as a robust version of the absolute $z$-score since
it uses robust measures of location and scale (instead of the usual 
mean and standard deviation). It also accounts for skewness by 
computing the scale on each side of the median separately.

As a robust measure of location we take the median of $Y$. 
The robust estimates of scale are based on the adjusted 
boxplot~\citep{Hubert:AdjBoxplot}. 
This modified boxplot is well-suited for skewed distributions. 
First, an interval called the fence is computed. 
For right-skewed data it is given by
\begin{equation*}
  [Q_1(Y)-1.5 \, e^{-4\text{MC}(Y)} \, \mbox{IQR}(Y) \; ,
   \; Q_3(Y)+1.5 \, e^{+3\text{MC}(Y)} \, \mbox{IQR}(Y)]
\end{equation*}
with $Q_1(Y)$ and $Q_3(Y)$ the first and third quartile of $Y$, 
$\mbox{IQR}(Y) = Q_3(Y)-Q_1(Y)$ and $\text{MC}(Y)$ the medcouple, 
a robust measure of skewness \citep{Brys:RobSkew}. 
For left-skewed data a similar definition holds. 
Note that for perfectly symmetric data the medcouple is zero, and 
then the fence is identical to that of the standard boxplot.
The whiskers of the adjusted boxplot are lines from the box to the 
most remote data points inside the fence.  

\begin{figure}[!htb]
\begin{center}
\unitlength=0.75mm \thicklines
\begin{picture}(170,50)
\put(5.00,20.00){\line(0,1){20.00}}
\put(5.00,30.00){\line(1,0){30.00}}
\put(35.00,20.00){\line(1,0){40.00}}
\put(35.00,20.00){\line(0,1){20.00}}
\put(35.00,40.00){\line(1,0){40.00}}
\put(75.00,20.00){\line(0,1){20.00}}
\put(75.00,30.00){\line(1,0){70.00}}
\put(145.00,20.00){\line(0,1){20.00}}
\put(50.00,20.00){\line(0,1){20.00}}
\put(0,30.00){\circle*{2}}
\put(5,30.00){\circle*{2}}
\put(100,30){\circle*{2}}
\put(160,30.00){\circle*{2}}
\put(145,30.00){\circle*{2}}
\put(3.00,25.00){\makebox(0,0)[rr]{$x_1$}}
\put(103.00,25.00){\makebox(0,0)[rr]{$x_2$}}
\put(8.00,44.00){\makebox(0,0)[rr]{$w_1$}}
\put(147.00,44.00){\makebox(0,0)[rr]{$w_2$}}
\put(25.00,16.00){\makebox(0,0)[rr]{$d_1$}}
\put(77.00,16.00){\makebox(0,0)[rr]{$d_2$}}
\put(25.00,5.00){\makebox(0,0)[rr]{$s_1$}}
\put(100.00,5.00){\makebox(0,0)[rr]{$s_2$}}
\put(54.00,44.00){\makebox(0,0)[rr]{$\med$}}
\put(38.00,44.00){\makebox(0,0)[rr]{$Q_1$}}
\put(77.00,44.00){\makebox(0,0)[rr]{$Q_3$}}

\put(10.00,8.00){\vector(1,0){40}}
\put(10.00,8.00){\vector(-1,0){5}}
\put(55.00,8.00){\vector(1,0){90}}
\put(55.00,8.00){\vector(-1,0){4}}
\put(5.00,13.00){\vector(1,0){45}}
\put(5.00,13.00){\vector(-1,0){5}}
\put(55.00,13.00){\vector(1,0){45}}
\put(55.00,13.00){\vector(-1,0){4}}
\end{picture}
\end{center}
\vspace{-0.5cm}
\caption{Adjusted outlyingness based on the adjusted boxplot.}
\label{fig:adjBox}
\end{figure}
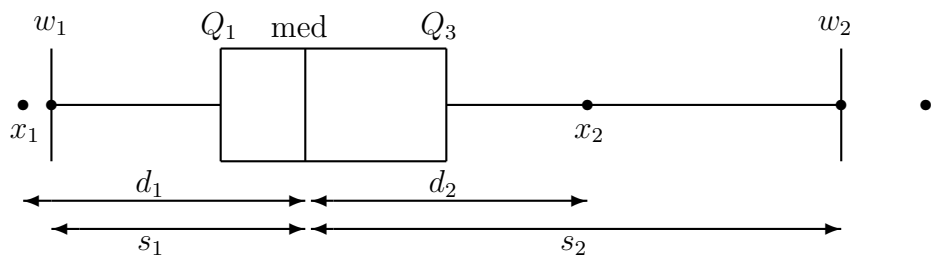

Figure~\ref{fig:adjBox} shows the adjusted boxplot of a 
right-skewed dataset $Y$. 
The points $x_1$ and $x_2$ have the same distance $d_1=d_2$ to the 
median of $Y$, but $x_1$ lies outside the lower whisker $w_1$ 
whereas $x_2$ lies inside the upper whisker. 
The adjusted outlyingness of $x_1 < \med(Y)$ is now defined as 
$d_1/s_1$ with $d_1 = \med(Y)-x_1$ and $s_1=\med(Y) - w_1$. 
For $x_2 > \med(Y)$ the AO is given by $d_2/s_2$ 
with $d_2=x_2-\med(Y)$ and $s_2 = w_2 - \med(Y)$. 
In this example $\AO(x_1) > \AO(x_2)$.\\

Formally, the univariate adjusted outlyingness AO of a point $x$ 
relative to a sample $Y$ is given by:

\begin{equation}\label{eq:adjOut_univ}
\mbox{AO}(x;Y)=
  \left\{
	\begin{array}{ll}
		\frac{x-\text{med}(Y)}{w_2(Y) - \text{med}(Y)}
		      & \;\; \mbox{ if } x > \med(Y) \\
		\frac{\text{med}(Y) - x}{\text{med}(Y) -  w_1(Y)}
		      & \;\; \mbox{ if } x < \med(Y) \;\;.
	\end{array} 
  \right. 
\end{equation}

To compute the AO of a $p$-dimensional point $x$ relative to 
an $n \times p$ data matrix $Y$, the following projection 
pursuit procedure is applied.
The point $x$ and all data points are projected on a direction $v$,
and we compute the univariate AO of the projected point $x'v$
relative to the projected dataset $Yv$  as in \eqref{eq:adjOut_univ}. 
This is repeated for many directions $v$, and the multivariate AO 
is defined as the maximum over all corresponding univariate AO’s:
\begin{equation}\label{eq:adjOut}
     \mbox{AO}(x;Y)=\sup_{||v||=1} \; \mbox{AO}(x'v;Yv) \;\;.
\end{equation}
This definition is based on the principle that a multivariate point 
is outlying with respect to a dataset if it is outlying in at least 
one direction of the multivariate space. 
Note that this outlying direction does not have to correspond to 
one of the coordinate axes. 

To compute the multivariate AO we have to rely on approximate 
algorithms, as it is impossible to consider the projections on 
{\it all} directions $v$ in $p$-dimensional space. 
\cite{Hubert:OutlierSkewed} suggest using $250p$ directions. 
A possible procedure to generate a direction is to randomly
draw $p$ data points, compute the hyperplane passing through
them, and then to take the direction orthogonal to it.
This approach guarantees that the multivariate AO is affine 
invariant. This means that the AO does not change when we add a
constant vector to the data, or multiply the data by a 
nonsingular $p \times p$ matrix.

\end{document}